\documentclass[pra,twocolumn,footinbib,superscriptaddress,showpacs]{revtex4-1}

\usepackage{amsmath,amsfonts,amssymb,bm}
\usepackage{dcolumn}
\usepackage[final]{graphicx}
\usepackage{bm}
\usepackage{comment}
\usepackage{color}

\newcommand{\be}{\begin{equation}}
\newcommand{\ee}{\end{equation}}
\newcommand{\bea}{\begin{eqnarray}}
\newcommand{\eea}{\end{eqnarray}}
\newcommand{\bit}{\begin{itemize}}
\newcommand{\eit}{\end{itemize}}

\newcommand{\bfl}{\begin{flushright}}
\newcommand{\efl}{\end{flushright}}

\newcommand{\ra}{\rangle}
\newcommand{\la}{\langle}

\begin{document}

\bibliographystyle{apsrev4-1}

\title{Mach-Zehnder interferometry with interacting trapped Bose-Einstein condensates }

\author{Julian Grond} 
\affiliation{Institut  f\"ur Physik, Karl--Franzens--Universit\"at Graz, 8010 Graz,  Austria}
\affiliation{Wolfgang Pauli Institut c/o Fak. Mathematik, Universit\"at Wien, 1090 Vienna, Austria}
\affiliation{Vienna Center for Quantum Science and Technology, Atominstitut, Technische Universit\"at Wien,  1020 Vienna, Austria}
\affiliation{Theoretische Chemie, Physikalisch--Chemisches Institut, Universit\"at Heidelberg, 69120 Heidelberg,  Germany}

\author{Ulrich Hohenester} \affiliation{Institut  f\"ur Physik, Karl--Franzens--Universit\"at Graz, 8010 Graz,  Austria}

\author{J\"org Schmiedmayer} \affiliation{Vienna Center for Quantum Science and Technology, Atominstitut, Technische Universit\"at Wien,  1020 Vienna, Austria}
  
\author{Augusto Smerzi}  \affiliation{  INO-CNR BEC Center and Dipartimento di Fisica, Universit\`a di Trento, 38123 Povo, Italy}

\begin{abstract}

We theoretically analyze a Mach-Zehnder interferometer with trapped condensates, and find that it is surprisingly stable against the nonlinearity induced by inter-particle interactions. The phase sensitivity, which we study for number squeezed input states, can overcome the shot noise limit and be increased up to the Heisenberg limit provided that a Bayesian or Maximum-Likelihood  phase estimation strategy is used. We finally demonstrate robustness of the Mach-Zehnder interferometer in presence of interactions against condensate oscillations and a realistic atom counting error.

\end{abstract}


% 03.75.Dg...Atom & neutron interferometry
% 06.20.Dk...Measurement & error theory
% 03.75.Gg...Entanglement & decoherence in BECs
% 37.25.+k...Atom interferometry techniques

\maketitle

Atom interferometry \cite{cronin:09} with trapped Bose-Einstein condensates (BECs) is a very promising tool for the most precise measurements.  The non-linearity of BECs makes it possible to create highly squeezed states, which should allow to surpass the classical shot noise limit for the phase sensitivity $\Delta\theta=1/\sqrt{N}$  by a factor of $\sqrt{N}$ up to the Heisenberg limit (HL) $\Delta\theta=1/N$ \cite{wineland:94,giovannetti:04}, where $N$ is the number of atoms in the condensates. 

Both Atom chips \cite{folman:02} and dipole traps \cite{grimm:00} allow for versatile control of trapped BECs, and coherent splitting and interference \cite{schumm:05,albiez:05} have been demonstrated. The preparation of moderately number squeezed states through splitting of a condensate by transforming a harmonic potential well into a double-well  \cite{javanainen:99} has been recently achieved experimentally \cite{esteve:08,maussang:10}, and it has been suggested to use optimal control strategies to create highly squeezed states at short time scales \cite{grond.pra:09} exploiting the atom-atom interactions. 

However, according to the current literature it is generally believed that interactions are detrimental for interferometry as they induce phase diffusion \cite{javanainen:97}, thereby decreasing the phase coherence \cite{esteve:08,gross:10,riedel:10,maussang:10,tikh:10,grond.NJP:10}. The proposed standard solution is making the interactions small by employing Feshbach resonances \cite{chin:10,roati:07} or using state selective potentials for internal degrees of freedom \cite{gross:10,riedel:10}.  This is not always possible, and in many cases not desirable, because Feshbach tuning requires field sensitive states which are on the other hand not ideal for precision interferometry. Moreover, residual interactions might still decrease the sensitivity.

In this paper, we analyze the \emph{Mach-Zehnder} (MZ) interferometer for BECs trapped in a double-well potential in presence of atom-atom interactions.  We show that the sensitivity is not substantially degraded by the interactions and Heisenberg scaling can be achieved with the resources of number squeezed input states and atom-number measurements as readout.  Our scheme is robust against mechanical excitations of the BEC and finite atom number detection efficiency.

\begin{figure}[h]
    \includegraphics[width=.79\columnwidth]{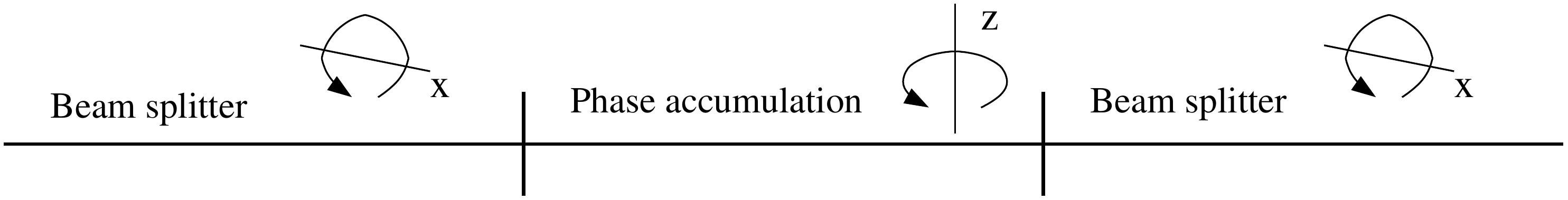} $\,\,\quad$\\
    \includegraphics[width=.92\columnwidth]{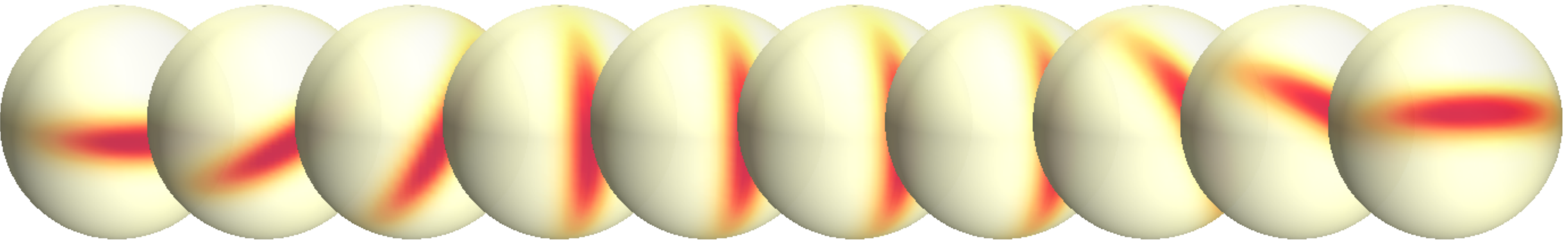} (a)\\
    \includegraphics[width=.92\columnwidth]{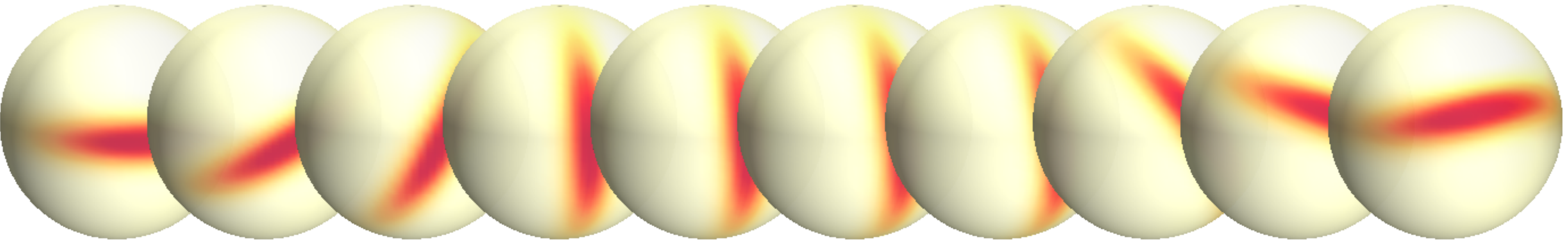} (b)\\
    \includegraphics[width=.92\columnwidth]{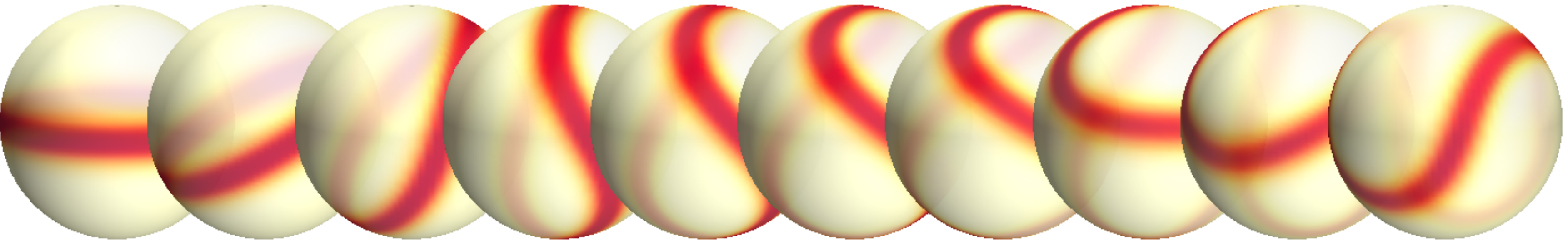} (c)
  \caption{(Color online) Mach-Zehnder interferometer sequence for a finite phase $\theta$ (a) in absence and (b-c) in presence of interactions, visualized on the Bloch sphere ($T_t=T_e=1$).  (a) A number squeezed initial state (small width along z-axis, number squeezing factor $\xi_N=0.2$) is transformed into a phase squeezed one (small width along equator) by a BS (rotation around $x$-axis). Next a phase is accumulated due to an external potential (rotation around $z$-axis). A second BS transforms the state such that the phase is mapped onto a number difference. (b) Even for very small interactions $U_0 N=0.1$ the number squeezing in the final state is lost. (c) For larger interactions $U_0 N=1$, the initial state (here a Fock state) gets strongly distorted and winds around the Bloch sphere. 
 \label{fig:MZBloch}}
\end{figure}

The initial state of the interferometer sequence consists of two uncoupled, stationary BECs with number fluctuations $\Delta J_z$ \cite{Note1}. We resort here to a generic description characterized by  two parameters  tunnel coupling $\Omega$ and interaction energy $U_0$ \cite{javanainen:99}, and discuss a realistic model at the end of the paper. We first introduce the \emph{ideal} (i.e., non-interacting) MZ interferometer as discussed in \cite{kitagawa:93,holland:93,pezze:09}. It consists of two \emph{cold atom beam splitters} (BS)  with Hamiltonian $H_{t}=-\Omega\hat J_x$, and in between a phase accumulation due to an energy difference $\Delta E$ between the two wells (with Hamiltonian $H_{e}=-\Delta E\hat J_z$). We visualize a typical interferometer sequence on the Bloch sphere \cite{grond.NJP:10} in Fig.~\ref{fig:MZBloch} (a). A BS corresponds to a $\pi/2$-rotation around the $x$-axis during a time $T_t$. The first BS transforms the number squeezed input state into a phase squeezed one. The second BS transforms an accumulated phase $\theta=\Delta E T_e$ (a $z$-rotation caused by an external potential during a time $T_e$) into a relative atom number difference. The whole interferometer transformation can be written as $|\psi_{\textrm OUT}^{(\theta)}\ra=e^{-i\theta\hat J_y}|\psi_{\textrm IN}\ra$. A number squeezed  input state [with squeezing factor $\xi_N:=\Delta J_z/(\sqrt{N}/2)<1$] reduces the measurement uncertainty in the atom number of the final state \cite{kitagawa:93,Note3}.    

% interferometer transformation

Atom-atom interactions are described by the Hamiltonian $H_{i}=U_0\hat J_z^2$ \cite{Note4}, and the whole interferometer transformation reads now
\be\label{eq:TFia}
|\psi_{\textrm OUT}^{(\theta)}\ra=e^{-i(H_{t}+H_{i})T_{t}}e^{-i(H_{e}+H_{i})T_{e}}e^{-i(H_{t}+H_{i})T_{t}}|\psi_{\textrm IN}\ra\,.
\ee
Even for very small interactions, $U_0 N=0.1$, the state gets distorted [Fig.~\ref{fig:MZBloch} (b)]. For larger interactions [Fig.~\ref{fig:MZBloch} (c)] the state covers almost the whole sphere. If we employ the usual parameter estimation based on the mean value of all the measurement results \cite{Note3,kitagawa:93,wineland:94}, the phase sensitivity is worse than shot noise.  

% QFI: define the problems to solve

Contrary to the expectations of this estimation, we will now show that interactions do not substantially limit interferometry. In a completely general fashion we use the \emph{Quantum Fisher information} $F_Q(|\psi_{\mathrm{IN}}\ra)=4(\Delta R)^2$, where $\hat R$ is the generator of the interferometer transformation \cite{holevo:82,pezze:09}, to compute the \emph{Cramer-Rao lower bound} (CRLB), which determines the best possible phase sensitivity independent of the choice of the measured observable \cite{Note5}. For the interferometer transformation Eq.~\eqref{eq:TFia} we find 
\be\label{eq:CRLB}
\Delta\theta_{\mathrm{CRLB}}\ge \frac{1}{\sqrt{m F_Q(|\psi_{\mathrm{IN}}\ra)}}=\frac{1}{\sqrt{m}2\Delta J_{z}(t=T_t)}\,,
\ee
i.e., it is given by the number fluctuations after the first nonlinear BS \cite{pezze2:06}. $m$ denotes the number of independent measurements. 

We start by analyzing a Fock input state $|\psi_{\mathrm{IN}}\ra \propto \bigl(\hat a_R^{\dagger}\bigr)^{N/2}\bigl(\hat a_L^{\dagger}\bigr)^{N/2}|0\ra$. From the scaling of $H=H_t+H_i$ with $N$ we find $\Delta J_z(t=T_t)\approx \alpha N $ (with constant $\alpha$). Thus, we expect Heisenberg scaling $\propto 1/N$ whenever  $U_0 N$ is constant for increasing $N$ \cite{Note6}. 

% CFI
Now we have to clarify whether one can indeed achieve a sensitivity close to the Heisenberg limit if one is restricted to a number measurement as in experiments. The \emph{Classical Fisher information} (CFI) \cite{holevo:82}
\be\label{eq:CFI}
F(\theta,|\psi_{\rm IN}\ra)=\int dn \frac{1}{P(n|\theta)}\Bigl(\frac{\partial P(n|\theta)}{\partial\theta}\Bigl)^2\,,
\ee
allows to estimate a lower bound of $\Delta \theta=1/\sqrt{m F(\theta)}$ for this specific type of measurement (we consider $\theta\ll 1$). Hereby, $P(n|\theta)=|\la n|\psi_{\rm OUT}^{(\theta)}\ra|^2$ is the conditional probability that an atom number difference $n$ is measured for phase $\theta$. Below we choose a constant $U_0 N=1$ and vary the BS and accumulation times $T_t$ and $T_e$ (with $\Omega T_t=\pi/2$ fixed). The influence of larger interactions can then be extracted through simple rescaling.

% CFI: Fock

\begin{figure}[h]
    \includegraphics[width=.99\columnwidth]{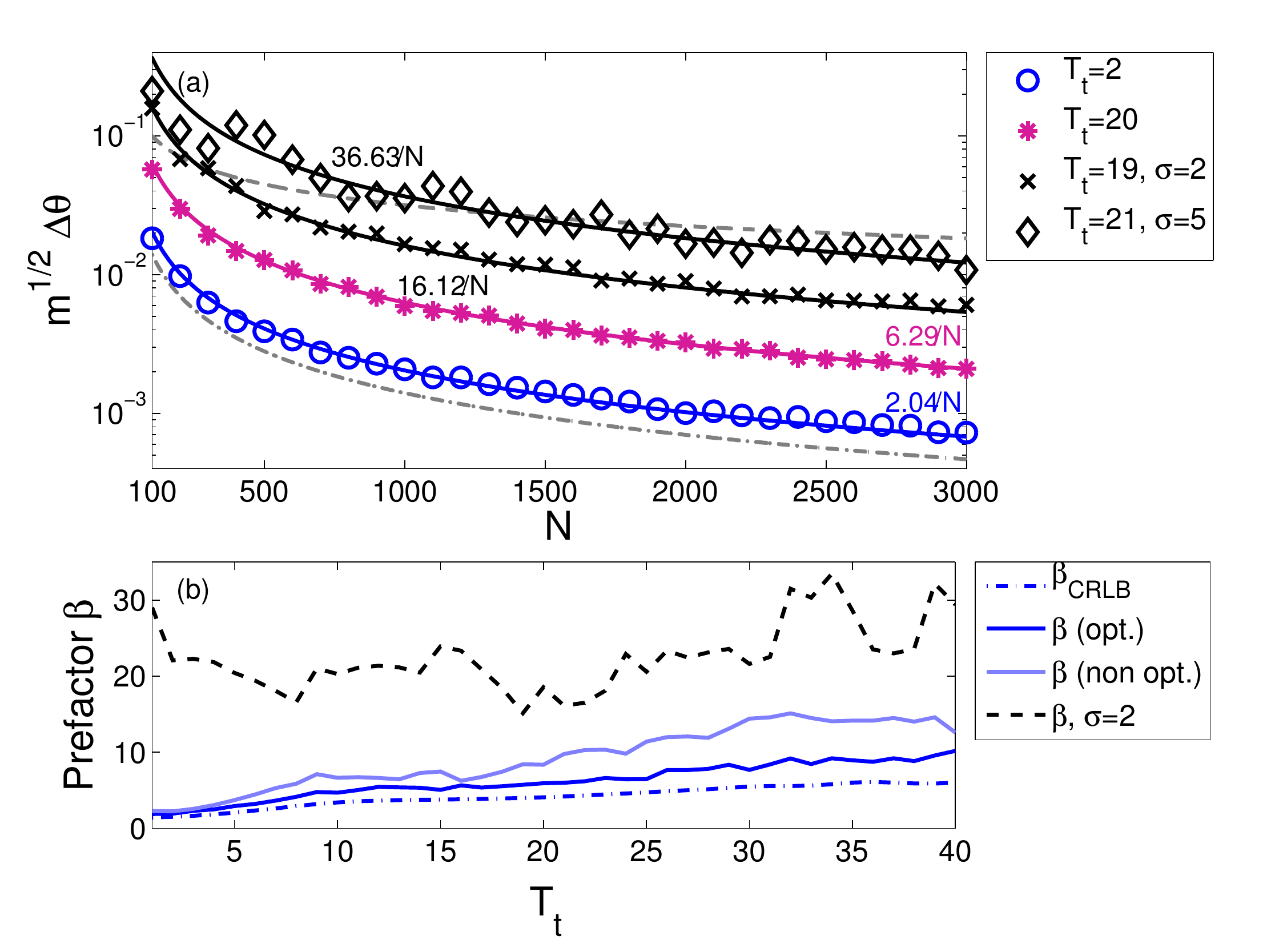}
  \caption{(Color online) (a) Scaling of $\sqrt{m}\Delta\theta$ with $N$ for $T_t=2$ (circles), $T_t=20$ (stars), and finite detection error $\pm 2$ (crosses, for $T_t=19$) and $\pm 5$ (diamonds, $T_t=21$)  ($U_0 N=1$ and $T_e$ fixed), compared to shot noise (dashed line) and the $U_0=0$ HL $\sqrt{m}\Delta\theta=1.4\sqrt{m}/N$ (dashed-dotted line).   (b) The prefactor $\beta$ (obtained from fitting) is shown for optimized $T_e<40$ (dark solid line), compared to $\sqrt{m}\Delta\theta_{\mathrm{CRLB}}\cdot N$ (dash-dotted line) and $\sqrt{m}\Delta\theta\cdot N$ for non-optimized values of $T_e<40$ (bright solid line). Also results for a finite detection error $\pm 2$ are shown (dashed line). \label{fig:ScaleCoeff}}
\end{figure}

Heisenberg scaling $\Delta\theta=\beta/N$ persists in presence of interactions also for a number measurement, see Fig.~\ref{fig:ScaleCoeff} (a).
The sensitivity is degraded only by an almost N-independent prefactor $\beta$, which varies with $T_t$ as is shown in Fig.~\ref{fig:ScaleCoeff} (b) (dark solid line).   Fast BSs [$T_t\sim 1/(U_0 N)$] give rise to a prefactor of $\sim 1$, but also for slower BSs we can exploit quantum correlations for MZ interferometry, which is relevant for relatively large interactions $U_0 N=10$ \cite{grond.NJP:10}. 

% substructures

The number readout works because interactions transform the conditional probability distributions $P(n|\theta)$, which have for the ideal MZ and $\theta=0$ a single peak with width $1$, 
\begin{figure}[h]
    \includegraphics[width=.99\columnwidth]{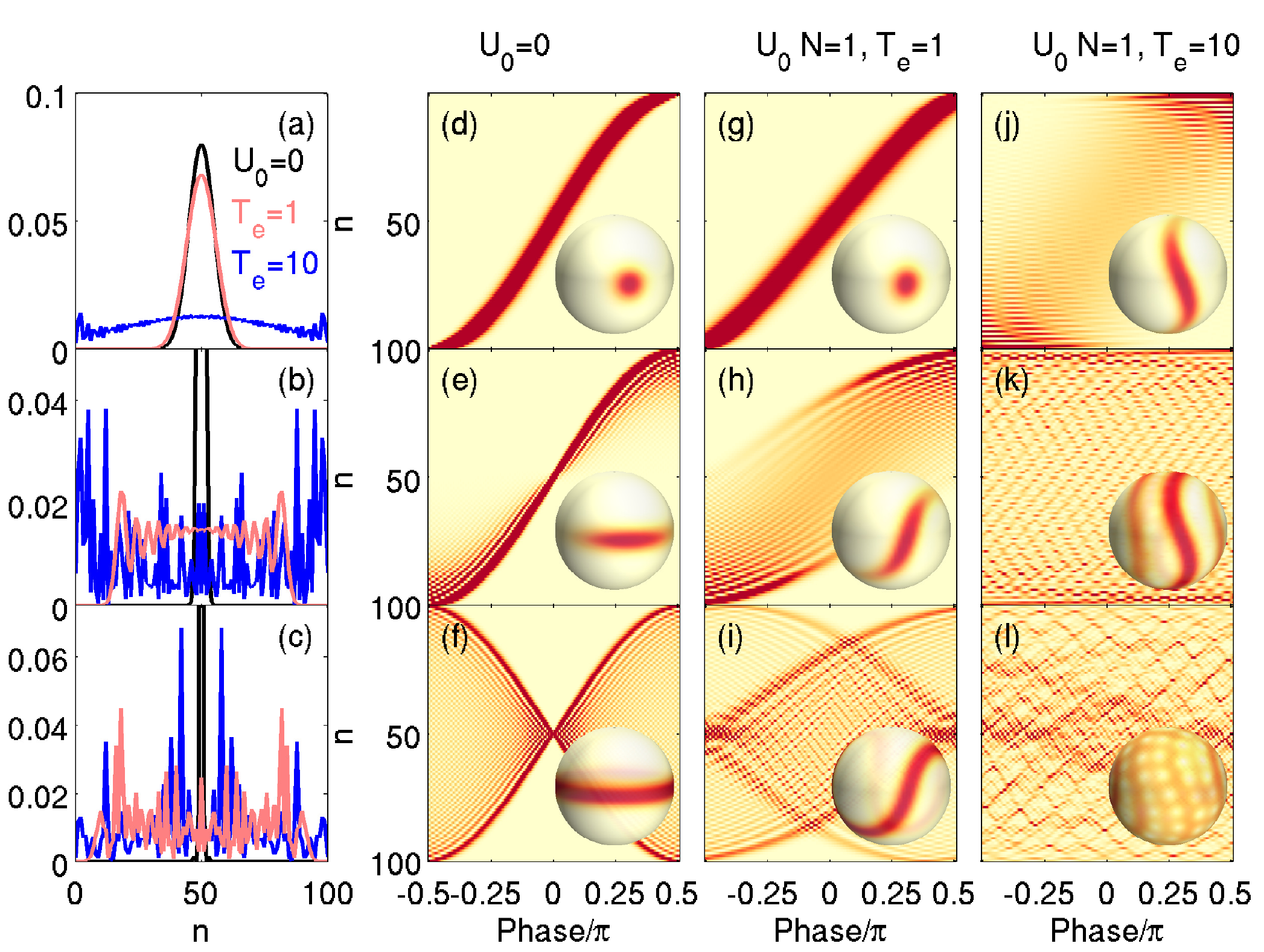}
  \caption{(Color online) Probabilities $P(n|\theta)=|\la n|\psi_{\rm OUT}^{(\theta)}\ra|^2$ for (a,d,g,j) binomial, (b,e,h,k) moderately number squeezed ($\xi_N=0.2$) and (c,f,i,l) Fock states for $N=100$, $U_0 N=1$ and $T_t=1$. The ideal case (no interactions) is shown in (d-f), the interacting case in (g-i) for $T_e=1$, and  (j-l) for $T_e=10$. The states are also shown on the Bloch sphere. (a-c) $P(n|\theta=0)$. \label{fig:Prob}} 
\end{figure}
into a complicated pattern with substructures of the same width [see Figs.~\ref{fig:Prob} (c,f,i,l)]. These serve as the measurement stick and determine the smallest phase which can be resolved \cite{pezze:09}. The patterns vary with $T_e$, such that some of them show up more distinct $1/N$-sized peaks [blue line in Fig.~\ref{fig:Prob} (c)], maximizing the CFI of Eq.~\eqref{eq:CFI} better than others (bright red line). 

Thus, the number measurement is not the 'optimal' measurement for all
values of $T_e$ \cite{Note5}. We compare the maximal and minimal prefactor which can
be obtained by varying $T_e$ [bright and dark solid lines in Fig.~\ref{fig:ScaleCoeff}
(b), respectively].  The latter lies close to the CRLB (dashed-dotted
line). Most importantly, there is no upper limit to $T_e$ which allows, in 
principle, to accumulate signals for a very long time.

% CFI: finite squeezing

\begin{figure}[h]
    \includegraphics[width=.99\columnwidth]{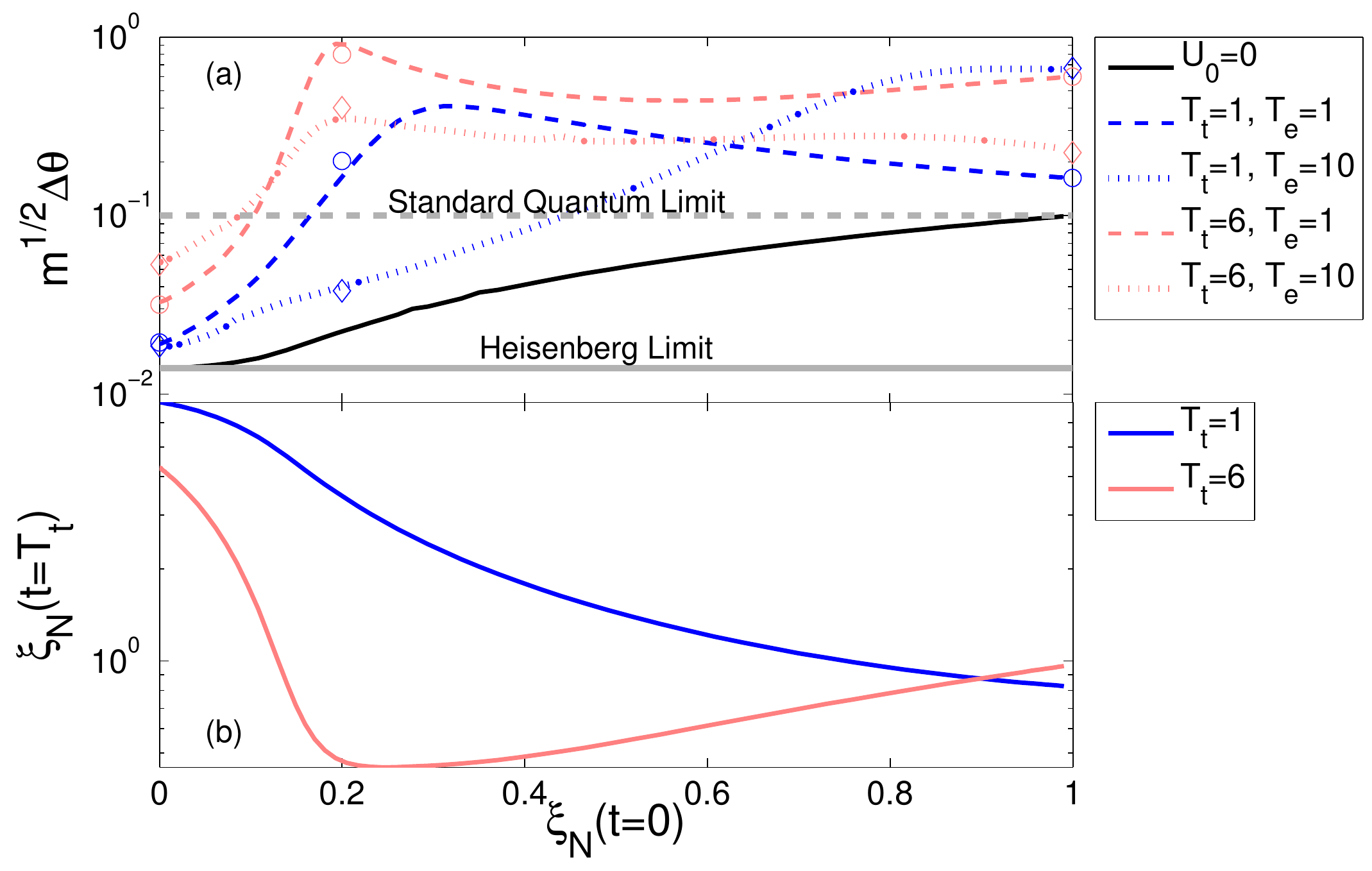}
  \caption{(Color online) (a) Phase sensitivity $\sqrt{m}\Delta\theta$ for different interaction strengths versus the initial number fluctuations $\xi_N(t=0)$ for $U_0 N=1$ ($N=100$). The symbols show results for a simulated Bayesian phase estimation (circles: $T_e=1$, diamonds: $T_e=10$). (b) Number fluctuations after the first BS, $\xi_N(t=T_t)$, which determine the CRLB Eq.~\eqref{eq:CRLB}.  \label{fig:FI}}
\end{figure}

In many experimental situations only input states with finite number squeezing $\xi_N<1$ are available. For the ideal MZ the sensitivity increases monotonously with number squeezing [black line in Fig.~\ref{fig:FI} (a)], up to the HL $\sqrt{m}\Delta\theta=1.4\sqrt{m}/N$.

We start with analyzing the case of \emph{long BS times} $T_t$ (bright red lines). We find a transition of the phase sensitivity as a function of $\xi_N$: Starting from $\xi_N=1$ (binomial state), the sensitivity first decreases up to a point, say around $\xi_N=0.2-0.3$. Then it becomes better again and finally approaches the HL for very small values of $\xi_N$. Also the CRLB, Eq.~\ref{eq:CRLB}, which is a strict lower bound to $\Delta\theta$, shows a transition. The reason is absence of number fluctuations after the first BS whenever the input state is only moderately number squeezed [bright red line in Fig.~\ref{fig:FI} (b)]. 

For \emph{short BS times} $T_t$ (blue lines) we find a transition only for short phase accumulation time $T_e$ (blue dashed line). In contrast, a longer $T_e$ gives a monotonous behaviour (blue dotted line) similar as the CRLB [blue line in Fig.~\ref{fig:FI} (b)].

We can get insight into this behaviour from the conditional probabilities $P(n|\theta)$ for different input states in Fig.~\ref{fig:Prob}. For binomial and moderately number squeezed states (d,e), they are close to Gaussian shape [black lines in (a-b)]. Interactions wash out the structure of the squeezed state (h) and increase  the variance in the final atom number distribution. Thereby the coherent spin squeezing  of the initial state is decreased  to values even worse as compared to the more robust binomial initial state \cite{tikh:10}. For longer $T_e$, interactions induce substructures (k).  In contrast, for a Fock input state a complex pattern emerges even for very small $T_e$ (i), whereas a binomial input state does not build up any substructure at all (g,j). Visualized on the Bloch sphere, $P(n|\theta)$ shows an interference pattern whenever a state winds around for a long enough time such that it becomes a superposition of different phase components [Figs.~\ref{fig:Prob} (k,i,l)].

%e$ (i).

In real experiments, $\Delta\theta$ as calculated from the CFI can be obtained by using a Bayesian (or alternatively Maximum-likelihood) phase estimation protocol \cite{pezze:07}. Thereby a series of $m$ measurements is performed, and the atom number difference of each measurement is used for the phase estimation. We find that such a protocol gives sensitivity in accordance with the  more general lower bounds as reported by the symbols in Fig.~\ref{fig:FI} (a) (for $m=20$). 

The MZ interferometer is robust against shot to shot fluctuations in the atom number or nonlinearity \cite{hyllus:10}. A finite atom counting error has the effect of broadening the substructures in the probability distributions as $\tilde{P}(n|\theta)\propto\sum_k P^{\mathrm{error}}(n|k) P(k|\theta)$, where $P^{\mathrm{error}}(n|k)$ is the error probability for measuring $k$ atoms instead of $n$. In Fig.~\ref{fig:ScaleCoeff} we show that a binomial error probability with width  $\sigma=2$ gives rise to just another prefactor, because a constant detection error is less important for larger N. Even for a detection error $\sigma=5$ \cite{esteve:08,ockeloen:10}, sub-shot noise can be found for  $N>2000$.  

\begin{figure}
\includegraphics[width=.95\columnwidth]{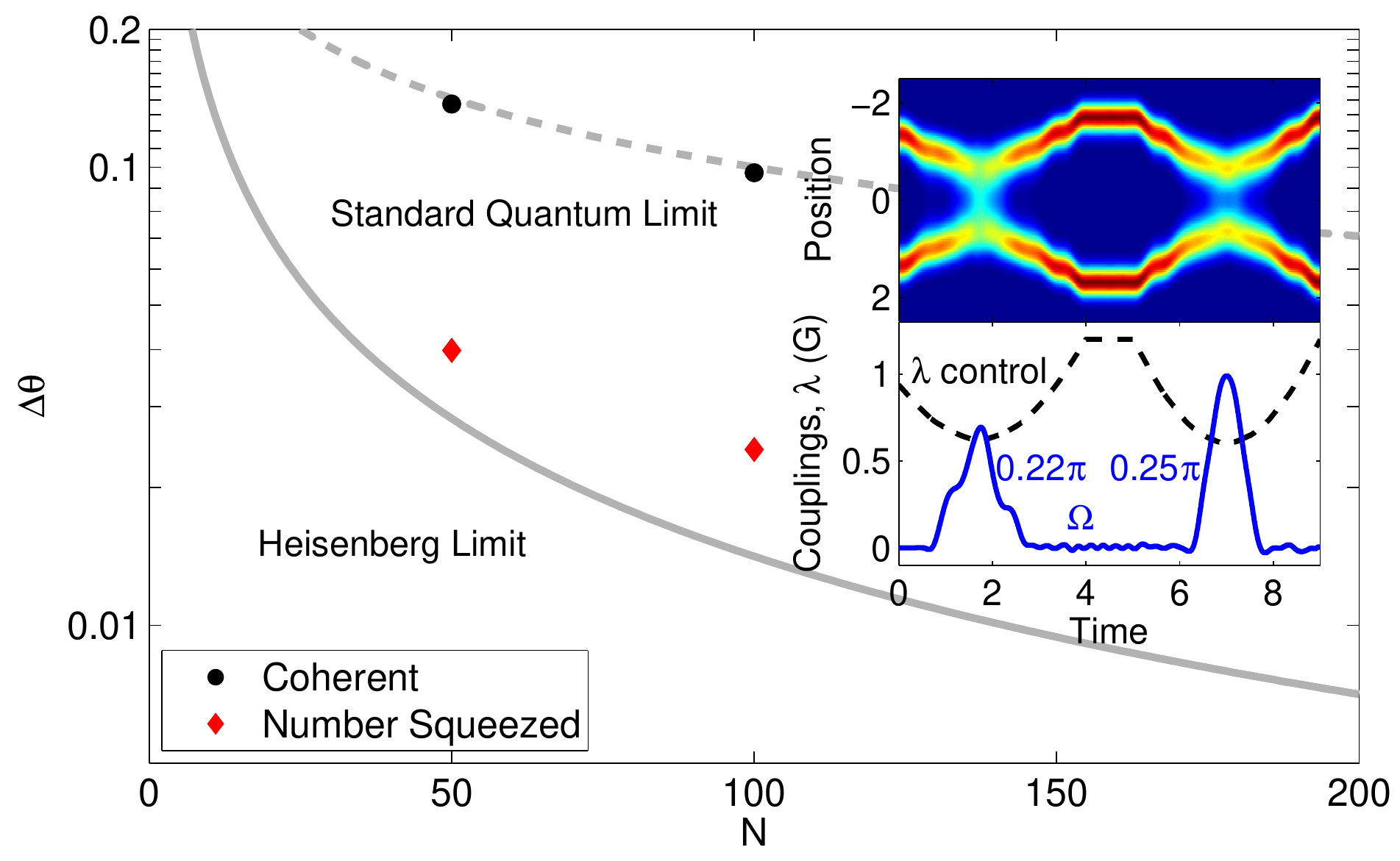}
  \caption{(Color online) $\Delta\theta$ for realistic control sequences calculated with MCTDHB for $U_0 N=0.1$ and a binomial state with $T_t=8$, and $U_0 N=1$ and a highly number squeezed state with $\xi_N(t=0)=0.05$ and $T_t=4$.  The insets show the  density (upper panel), the optimal control and tunnel coupling (lower panel) for the squeezed state and $N=100$. \label{fig:mzMCTDHB}}
\end{figure}

Implementing the interferometer with trapped condensates, one achieves the BS by lowering the barrier between two split condensates, thereby introducing tunneling. The full two-mode physics including the spatial dynamics can be accounted for by the \emph{Multi-configurational time-dependent Hartree for bosons} (MCTDHB) method \cite{alon:08}, which represents a framework using time-dependent mode functions.  For a typical trapping geometry on atom chips with $\omega_{\perp}=2\pi\cdot2$ kHz transverse frequency, we find that for tunnel pulses on the order of several milliseconds, rapid oscillations are induced in the condensates, which lead to unwanted excitations \cite{negretti:04}.   %On the other hand, a tunnel pulse with purely adiabatic motional dynamics is achievable only on the order of tens of $ms$, which leads to a severe decrease in phase sensitivity. 
In our earlier work \cite{grond.pra:09,grond.NJP:10} we have developed and demonstrated optimal control \cite{peirce:88,treutlein.pra:06} within MCTDHB. This allows us to design controls for fast BS operations without exciting the condensates, which is achieved by \emph{trapping} the condensates in stationary states after each of the two BSs, while at the same time achieving appropriate tunnel pulses. An approximately $\pi/4$-tunnel pulse \cite{Note7} is achieved for $T_t=4$ and highly number squeezed input states. In Fig.~\ref{fig:mzMCTDHB} we show results for binomial and number squeezed input states with a phase sensitivity close to the HL.   

To summarize, we analyzed the phase sensitivity of a trapped BEC Mach-Zehnder interferometer in presence of interactions. Heisenberg scaling can be achieved for an atom number measurement, and there is no upper limit to the phase accumulation time. For finitely number squeezed input states the phase sensitivity is characterized by a transition.    We demonstrated robustness against condensate oscillations and finite detection error, and thus our results can be compared to current experiments.

We thank F. Piazza, J. Chwede\'nczuk, T. Schumm, and S. Whitlock for most helpful discussions. JG is thankful for the great hospitality experienced during his visits in Trento, and acknowledges support from the Alexander von Humboldt-foundation,. This work has been supported in part by NAWI GASS, the FWF,  and the ESF Euroscores program: EuroQuaser project QuDeGPM.

\end{document}